\documentclass[onecolumn,11pt]{article}
\usepackage{amsmath}
\usepackage{amsfonts}
\usepackage{amssymb}

\newcommand{\ket}[1]{|#1\rangle}

\newcommand{\absq}[1]{|#1|^2}
\newcommand{\boxit}[1]{\vskip6pt\vbox{\hrule\hbox{\vrule\kern8pt\vbox{\kern8pt\vbox{\hsize344pt\noindent\strut#1\strut}\kern8pt}\kern8pt\vrule}\hrule}}
\title{\textbf{Interpreting Bananaworld: A response to \\ Bub's Quantum Mechanics for Primates}}
\author{Ulrich Mohrhoff\\
\textit{\small Sri Aurobindo International Centre of Education}\\ 
\textit{\small Pondicherry 605002, India}\\
\ttfamily{\small ujm@auromail.net}}
\date{}
\normalsize
\begin{document}
\maketitle
\bigskip
\begin{abstract}
\noindent The interpretative principle proposed by Bub in 1211.3062v1 [quant-ph] is justified only for all practical purposes (Bell's ``FAPP trap''). An alternative interpretative principle is proposed. It brings to light those features of the quantum world because of which the fundamental theoretical framework of physics is a ``mere'' probability calculus, and it amply justifies Bohr's insistence that quantum-mechanical observables cannot be defined without reference to the experimental conditions in which they are measured. It implies that the spatial distinctions we make cannot be intrinsic to space, that regions ``of space'' must be realized by macroscopic objects, that the spatiotemporal differentiation of the physical world is incomplete, that the positions of macroscopic objects (suitably defined) are definite in a nonclassical sense, and that unconditional factuality can be consistently attributed to them.
\end{abstract}
\bigskip
\section{Introduction}
In his recent paper ``Bananaworld: Quantum Mechanics for Primates''~\cite{BubBW}, Jeffrey Bub writes (and I could not agree more) that
\begin{quote}
quantum mechanics is about probabilistic correlations, i.e., \emph{about the structure of information}, insofar as a theory of information is essentially a theory of probabilistic correlations---not about energy being quantized in discrete lumps or quanta, not about particles being wavelike, not about the universe continually splitting into countless co-existing quasi-classical universes, with many copies of ourselves, or anything like that.	[original emphasis]
\end{quote}
Having said this, and taking for granted that ``the dynamics\dots generally produces entanglement between two coupled systems,'' Bub raises the question whether ``the dynamics\dots is consistent with the assumption that something definite happens in a measurement process'':
\begin{quote}
The basic question is whether it is consistent with the unitary dynamics to take the macroscopic measurement ``pointer'' or, in general, the macroworld as definite.
\end{quote}
To my mind, the fact that quantum mechanics is about probabilistic correlations implies that quantum mechanics is a probability calculus. The events to which this assigns probabilities, as well as the data on the basis of which the probabilities are assigned, are measurement outcomes. This being so, I cannot but wonder how quantum mechanics can at the same time be about (let alone describe) a linear, unitary dynamics that ``generally produces entanglement between coupled systems.'' I can't seem to reconcile these notions. Rather, I think that Bub's list of what quantum mechanics is not about lacks a crucial item: in addition to not being about energy being quantized in discrete lumps or quanta, about particles being wavelike, about the universe continually splitting into countless co-existing quasi-classical universes, \emph{it is not about a dynamical process of any sort}.

Bub further writes (and again I could not agree more) that ``we should not expect to find a dynamical explanation for the selection of a particular outcome in a quantum measurement'':
\begin{quote}
The selection of a particular outcome in a quantum measurement, and the associated loss of complementary information, is a genuinely random event, a ``free choice'' on the part of the system, not the culmination of a dynamical process.
\end{quote}
Since the quantum-mechanical probability calculus presupposes the events to which, and on the basis of which, it serves to assign probabilities, it should not come as a surprise that quantum mechanics cannot account for the existence of these events~\cite{UlfbeckBohr01,Mohrhoff-clicks}. Quantum-mechanical probabilities are conditional probabilities, and this for at least two reasons. One is that they are assigned to possible measurement outcomes \emph{on the basis of actual outcomes}. This is a direct consequence of the fact that quantum mechanics is about probabilistic correlations. If the possible outcomes of two measurements are correlated, the actual outcome of either measurement can be used as the basis for assigning probabilities to the possible outcomes of the other measurement. (This is as true for measurements performed on the same system at different times as it is for measurements performed on different systems in spacelike relation.)

Another reason why quantum-mechanical probabilities are conditional is that they are assigned to possible measurement outcomes \emph{conditional on the existence of an outcome}. (This after all is why the probabilities of the possible outcomes of a measurement add up to~1.) The probability $P(q,Q,t)$ with which a measurement of $Q$ \emph{attempted} at the time $t$ yields the outcome $q$ is the product of two probabilities: the probability $P(Q,t)$ that the measurement \emph{succeeds} (which, as every experimental physicist can confirm, is by no means certain), and the probability $p(q|Q,t)$ that the outcome of a measurement of~$Q$, attempted at the time~$t$, is $q$ just in case the measurement succeeds:
\[
P(q,Q,t)= p(q|Q,t)\,P(Q,t).
\]
Quantum mechanics is exclusively concerned with conditional probabilities of the form $p(q|Q,t)$. Here again there are two conditions: $Q$ and~$t$. The fact that $Q$ appears as a condition tells us that quantum mechanics is incapable of predicting whether there will be an outcome. (This must not be confused with the theory's inability to predict the actual outcome whenever probabilities greater than~0 are assigned to two or more possible outcomes.) Because quantum mechanics ``knows nothing'' about causally sufficient conditions for (successful) measurements, the appearance of an actual outcome cannot be anything other than what Bub declares it to be, ``a truly random event.''

The fact that $t$ appears as a condition, on the other hand, tells us that the time on which a quantum-mechanical probability algorithm---a state vector, a wave function, or a density operator---functionally depends is not the continuous time-dependence of an evolving physical state. Rather, it is the time of the measurement to the possible outcomes of which the algorithm serves to assign probabilities.

What, then, could be the sense in which the ``dynamics\dots generally produces entanglement between two coupled systems''? Suppose that the probability algorithm
\[
\ket{\Psi(t_1)} = \left(\sum_i c_i\,\ket{a_i(t_1)}\right)\otimes\ket{b_0(t_1)}
\]
is associated with a composite system. It tells us that a (successful) joint measurement performed on the component systems at the time $t_1$ will yield the pair of values $a_i,b_0$ with probability $\absq{c_i}$. Suppose next that the component systems interact between the times $t_2>t_1$ and $t_3>t_2$, and that the probability algorithm \emph{for} (not \emph{at}) a time $t_4>t_3$ is
\[
\ket{\Psi(t_4)} = \sum_i c_i\,\ket{a_i(t_4)}\otimes\ket{b_i(t_4)}.
\]
This tells us that a (successful) joint measurement performed on the component systems at the time $t_4$ will yield the pair of values $a_i,b_i$ with probability $\absq{c_i}$. The ``dynamics'' which ``produces'' the entangled state $\ket{\Psi(t_4)}$ does not consist in the continuous evolution of a physical state $\ket{\Psi(t)}$ from $t_1$ to $t_4$. Rather, it consists in the substitution of one probability algorithm for another. $\ket{\Psi(t)}$ is not an instantaneous state that exists at the time~$t$, and this for the same reason that the probability for something to happen at the time $t$ is not something that exists at the time~$t$.

But does not the change from $\ket{\Psi(t_2)}$ to $\ket{\Psi(t_3)}$ describe the interaction that takes place between $t_2$ and $t_3$? And if so, does it not describe a continuous process? Because the answer to the first question is negative, the second question does not arise. The change from $\ket{\Psi(t_2)}$ to $\ket{\Psi(t_3)}$ describes the \emph{effect} this interaction has \emph{on probability assignments} to the possible outcomes of whichever joint measurement is performed after~$t_3$.

Finally suppose that the probability algorithm for the time $t_4$ is
\[
\ket{\Psi(t_4)} = \sum_i c_i\,\ket{e_i(t_4)}\otimes\ket{a_i(t_4)}\otimes\ket{b_i(t_4)},
\]
where the kets $\ket{b_i}$, $\ket{a_i}$, and $\ket{e_i}$ are states of a system, an apparatus, and the (internal and external) apparatus environment, respectively. The environment will be so strongly correlated with the rest of the world that it makes little sense to distinguish between the environment and the rest of the world. In other words, the environment \emph{is} the rest of the world (relative to system and apparatus). $\ket{\Psi(t_4)}$ now tells us that a joint measurement performed on the system, the apparatus, and the rest of the world at the time $t_4$ will yield the respective values $e_i,a_i,b_i$ with probability $\absq{c_i}$. Since quantum-mechanical probability assignments are conditional on the successful performance of a measurement of a specified observable at a specified time, and since it is logically impossible to subject the rest of the world to a measurement, all that \emph{this} ``entangled state at the end of a quantum measurement process'' does is to reduce to absurdity the unitary dynamics as Bub understands it.

Bub's basic question---``whether it is consistent with the unitary dynamics to take the macroscopic measurement `pointer' or, in general, the macroworld as definite''---thus is a no-brainer. Quantum mechanics is about probabilistic correlations. What happens in a measurement is not a transfer of indefiniteness from the observable measured to the pointer reading but the reverse: a transfer of definiteness (or factuality) from the pointer reading to the observable measured. If a unitary dynamics ``takes place,'' it takes place \emph{between} the correlata whose correlations it encapsulates. There is no mystical arena, unobservable to mortals, which evolves according to a unitary dynamics, and in which matters of fact about the values of observables pop up at random. There are (i)~matters of fact about the values of observables and (ii)~the unitary transformations by which they are probabilistically correlated. (It may be wholesome to bear in mind that the ``dynamics'' is unitary only because the probabilities of the possible outcomes of a measurement are supposed to add up to~1.)  

The basic question is not how the unitary dynamics can be consistent with a definite macroworld. The basic question is how quantum mechanics can be at once a fundamental physical theory and a ``mere'' probability calculus, concerned with nothing but probabilistic correlations. As a probability calculus, it presupposes the value-indicating facts to which (and on the basis of which) probabilities are assigned. It presupposes an actual world containing events that are governed by irreducibly probabilistic correlation laws. As a fundamental physical theory, it must make it possible to identify a set of observables whose values are factual \emph{per se} (as opposed to factual only because they are indicated by pointers whose positions are factual \emph{per se}).

The emphasis is on factuality---for which Bub uses Einstein's expression \emph{So-Sein} (``being thus'')---rather than on definiteness. The two concepts are by no means equivalent. Definiteness is sufficient for factuality. Because the classical world is definite, we may think of it as factual \emph{per se}.%
\footnote{The state of a classical system with $n$ degrees of freedom can be represented by a point $\cal P$ in a $2n$-dimensional phase space~$\cal S$, and the positive outcome of an elementary test can be represented by a subset ${\cal U}\subset {\cal S}$. The probability of obtaining $\cal U$ is equal to unity if ${\cal P} \in {\cal U}$ and equal to zero if ${\cal P}\notin {\cal U}$. Because the probability algorithm $\cal P$ only assigns the trivial probabilities 0 and 1, it may be thought of as a state in the classical sense of the word: a collection of possessed properties. We are at liberty to believe that the probability of finding the property represented by $\cal U$ is equal to unity \emph{because} the system has it.}
The converse is not true. Factuality is not sufficient for definiteness, unless it be a nonclassical definiteness (see Sec.~\ref{sec:macroworld}). The good news here is that the basic question can be answered without demonstrating the consistency of attributing definiteness to the macroworld. Demonstrating the consistency of attributing factuality to the macroworld will suffice.

The identification of an observable $R$ whose value is (or can  be consistently stipulated to be) either definite or factual \emph{per se} calls for an interpretative principle. Bub proposes to treat the decoherence pointer selected by environmental decoherence as such an observable. In Sec.~\ref{sec:bip} I explain why Bub's interpretative principle justifies the attribution of definiteness to the pointers selected by environmental decoherence only \emph{for all practical purposes}. In Sec.~\ref{sec:yaip} I propose an interpretative principle that (i)~implies the strict consistency of attributing factuality to the macroworld, (ii)~brings to light those features of the quantum world because of which the fundamental theoretical framework of physics is a ``mere'' probability calculus, and (iii)~amply justifies Bohr's insistence that quantum-mechanical observables cannot be defined without reference to the experimental conditions in which they are measured. The final task---to define the macroworld in terms of the quantum-mechanical probability calculus and to demonstrate the strict consistency of attributing to it an unconditional factuality---is carried out in Sec.~\ref{sec:macroworld}.

\section{Bub's interpretative principle}
\label{sec:bip}
Here is how Bub states (and justifies) his interpretative principle:
\begin{quote}
If we take $R$ as the decoherence ``pointer'' selected by environmental decoherence, then it follows that the macroworld is always definite because of the nature of the decoherence interaction coupling environmental degrees of freedom to macroworld degrees of freedom.  
\end{quote}
In an earlier paper~\cite{BubQC} Bub argued that 
\begin{quote}
the inconsistency in reconciling the entangled state at the end of a quantum measurement process with the definiteness of the macroscopic pointer reading and the definiteness of the correlated value of the measured micro-observable is only apparent and depends on a stipulation that is not required by the structure of the quantum possibility space. Replacing this stipulation by an alternative consistent stipulation resolves the problem.
\end{quote}
The stipulation that is not required by the quantum-mechanical probability calculus is an interpretative principle sometimes referred to as the ``eigenvalue-eigenstate link,'' which Dirac~\cite[pp. 46--47]{Dirac} formulated as follows:
\begin{quote}
The expression that an observable ``has a particular value'' for a particular state is permissible in quantum mechanics in the special case when a measurement of the observable is certain to lead to the particular value, so that the state is an eigenstate of the observable.
\end{quote}
(The reason why I likewise reject the eigenvalue-eigenstate link is not that it makes it impossible to reconcile the entangled state ``at the end of a quantum measurement process'' with the definiteness of the macroscopic pointer reading but that $\ket{\Psi(t)}$ is not an instantaneous state existing at the time~$t$, anymore than the probability that something happens at the time $t$ exists at the time~$t$.) The alternative stipulation, which Bub claims resolves the problem, is to attribute definiteness to the pointer selected by environmental decoherence.

Detailed investigations of a large number of specific models have demonstrated that the reduced density operators associated with sufficiently large and/or massive systems, obtained by partial tracing over ``realistic'' environments, become very nearly diagonal with respect to a privileged pointer basis in a very short time, and that they stay that way for a very long time \cite{Zurek2003, Blanchardetal2000, Joosetal2003, Schlosshauer04, Schlosshauer07}. ``Very nearly diagonal,'' however, is not the same as ``diagonal.'' It is equivalent to ``diagonal FAPP''~\cite{Bell1990}. 

If the off-diagonal (or interference) terms would vanish completely and for all time, the selection (by us) of a single diagonal element would be as unproblematic as the selection (by us) of the actual world in the context of classical physics. (Physical theories deal with nomologically possible worlds. No physical theory, whether classical or quantum, can distinguish the actual world from the other nomologically possible worlds.%
\footnote{This is another reason why quantum theory ought not to be called upon to distinguish the actual outcome of a measurement from the other possible outcomes.}
One salient difference between classical physics and quantum physics is that the former does not cause anybody to worry about how possibilities become facts.) Because the off-diagonal terms are not strictly~0, Bell's criticism (``FAPP-trap!'') also applies to Bub's resolution of the problem. Because the interference terms are extremely small, it is impossible \emph{for all practical purposes} to observe the coherence in superpositions of pointer states, and this justifies the attribution of definiteness to the pointers selected by environmental decoherence \emph{for all practical purposes}.

\section{Yet another interpretative principle}
\label{sec:yaip}
More can be accomplished. I will now formulate an interpretative principle that
\begin{enumerate}
\item implies the strict consistency (rather than the consistency FAPP) of attributing factuality (rather than definiteness) to the macroworld,
\item brings to light those features of the quantum world because of which the fundamental theoretical framework of physics is a ``mere'' probability calculus,
\item and amply justifies Bohr's insistence that quantum-mechanical observables cannot be defined without reference to the experimental arrangements by which they are measured.%
\footnote{See for instance Ref.~\cite{Bohr63}, wherein Bohr states that ``the limited commutability of the symbols by which [kinematical and dynamical] variables are represented in the quantal formalism corresponds to the mutual exclusion of the experimental arrangements required for their unambiguous definition.'' D'Espagnat~\cite{dEspagnat76} formulates one of ``the two pillars on which Bohr's doctrine is built'' in the following way: ``In general, quantum systems should not even be thought of as possessing individual properties [such as position and momentum] independently of the experimental arrangements.''}
\end{enumerate}
If a quantum state's dependence on time were the continuous time dependence of an evolving physical state, quantum states would be instantaneous states evolving according to a unitary dynamics, except (possibly) at the time of a measurement, when they are said to collapse or appear to do so. Having disposed of the measurement problem in its crudest form---Why two modes of evolution for the price of one?---we are left with two Lorentz invariant computational rules, one requiring us to first square the absolute values of amplitudes and then add the result, the other requiring us to first add amplitudes and then square the absolute value of the result. Why two? What is the essential difference between their respective conditions of applicability?

Here is the interpretative principle I propose: 

\boxit{Whenever quantum mechanics instructs us to first add the amplitudes associated with alternatives and then square the absolute value of the result, \emph{the distinctions we make between the alternatives correspond to nothing in the real world}. They exist solely in our minds.}

All it now takes to unearth the single most important feature of the quantum world, is the two-slit experiment with electrons. (When Feynman said that this experiment ``has in it the heart of quantum mechanics''~\cite{Feynman65}, he may have been more right than he knew.) If quantum mechanics instructs us to add amplitudes, it is not the case that the individual electron goes through \emph{either} the left slit \emph{or} the right slit. Somehow an electron can go through both slits (as a whole, without being divided into parts%
\footnote{The question of parts does not arise. Analogous experiments have been performed with C$_{60}$ molecules using a grating with 50-nm-wide slits and a 100-nm period~\cite{Arndtetal99}. We do not picture the individual carbon nuclei as getting separated by many times 100 nanometers and then reassemble into a ball less than a nanometer across. Nor is a probability algorithm like the wave function something possessing parts that pass through different slits.}
that go through different slits).

But how is it possible for a particle to go through both slits? This would indeed be impossible if the two slits were different parts of space, which is what we are inclined to think. Let us then ask: how exactly do the two slits in a double-slit experiment differ? They are cutouts in a slit plate---things that have been removed, things that are no longer there. What difference do they leave behind after they have been removed? The difference between the positions they previously occupied? But positions are properties (or relational properties, if you prefer, or properties of pairs), and properties exist only if and while they are possessed. Or do they?

There is something peculiar about the way we tend to think about space. We all more or less readily agree that red, round, or a smile cannot exist without a red or round object or a smiling face. This is why the Cheshire cat strikes us as funny. Why then do we tend to believe that positions exist by themselves, without being possessed?

It has not always been so. Influential thinkers from Aristotle to  Kant and Gauss have insisted that \emph{potential} infinities, such as the \emph{possibility} of conceptually dividing space ad infinitum, should be thought of as just that---possibilities rather than actualities. Kant~\cite{Kant1929} wrote that the so-called parts of space
\begin{quote}
cannot precede the one all-embracing space, as being, as it were, constituents out of which it can be composed; on the contrary, they can be thought only as in it.
\end{quote}
What Kant says about the parts of space applies, a fortiori, to the so-called points of space. In the second half of the nineteenth century mathematics nevertheless shifted to dealing with the continuum as a set of points. ``So successful has this shift been,'' von Weizs\"acker~\cite[p.~347]{vW1980} remarked, ``that it is nearly impossible to disabuse the contemporary student of mathematics of the superstition that this conception is the only possible, indeed `the' theory of \dots `the' continuum.''

If one calls a self-adjoint operator ``elephant'' and a spectral decomposition ``trunk,'' one can prove that every elephant has a trunk. Likewise, if one calls a real number ``point'' and the transfinite manifold of real numbers ``continuum,'' one can think of the continuum as composed of points. But does this mathematical continuum have more in common with physical space than the spectral theorem has with certain pachyderms? Von Weizs\"acker~\cite[p.~130]{vW1980} did not think so:
\begin{quote}
The conception of the continuum as potential, which originated with Aristotle, appears to be more suitable for the quantum theoretical way of thinking than is the set-theoretical conception of an actually existing transfinite manifold of ``real numbers,'' or of the spatial points they designate. The ``real number'' is a free creation of the human mind and perhaps not conformable to reality.
\end{quote}
If proof is needed that the set-theoretic conception of space is not conformable to reality, it is the ability of a particle to go through more slits than one (as a whole, without being divided into parts that go through different slits). 

The reason why it is possible for a particle to go through both slits is that the distinction we make between the spatial alternatives defined by the slits has no reality as far as the electron is concerned. 

But if there is a single physical system for which this distinction has no reality, then it cannot be intrinsic to space. What, then, furnishes space with its so-called parts? The answer: macroscopic detectors.%
\footnote{Such detectors are found not only in physics laboratories. Any macroscopic object equipped with a sensitive region $R$ and capable of indicating the presence in $R$ of another physical object qualifies.}
By \emph{realizing} (making real) a specific region ``of space,'' a detector makes it possible to attribute to another object the property of being in that region. And this bears generalization: \emph{the measurement apparatus is needed not only to indicate the answer to a question but also, and in the first place, to define a question by making its possible answers available for attribution.} This is the reason \emph{why the events to which quantum mechanics assigns probabilities are outcomes of measurements}.

But if it is impossible to attribute to a physical object the property of being in a region ``of space'' unless this is realized by a detector, then the spatiotemporal differentiation of the physical world cannot be complete---it cannot go ``all the way down.'' Because the uncertainty principle prevents physical objects from having definite positions (relative to each other), it is impossible for a detector to realize a definite position. This allows us to conceive of a partition of space into \emph{finite} regions so small that none of them is realized. But if none of them is realized, the spatial differentiation of the physical world is incomplete---it does not go ``all the way down.''

The same applies to the world's temporal differentiation, and this not merely because of the relativistic interdependence of distances and durations. Just as the properties of quantum systems or the values of quantum observables need to be realized by macroscopic devices, so the times at which properties or values are possessed need to be realized by macroscopic clocks. And just as it is physically impossible for macroscopic devices to realize sharp positions, so it is physically impossible for macroscopic clocks to realize sharp times~\cite{Hilgevoord98}.

But if neither the spatial nor the temporal differentiation of the physical world goes ``all the way down,'' determinism is out, and so is that mystical arena in which quantum states evolve and outcome-indicating facts pop up at random.%
\footnote{Bohr was right, there is no quantum world \emph{in this particular sense}, Bub's politically correct jibe at ``Copenhagen or neo-Copenhagen instrumentalism'' notwithstanding. (One may disagree with Bohr about what there is instead.)}
And if the dynamics (literally construed) is out, there can be no kind of evolving instantaneous physical state (inasmuch as such a state, existing as it does at every instant of time, cannot exist in the absence of a completely differentiated time), and all that can be predicted is probabilities. This is the reason \emph{why the fundamental theoretical framework of physics is a probability calculus}.

\section{The macroworld}
\label{sec:macroworld}
One task remains: to define the macroworld in terms of the quantum-mechanical probability calculus and to demonstrate the strict consistency (rather than the consistency FAPP) of attributing to it an unconditional factuality.

The possibility of obtaining factual evidence of the departure of an object $O$ from its classically predictable position calls for detectors whose position probability distributions are narrower than $O$'s. (By the classically predictable position of $O$ I mean the position that can be predicted on the basis of all relevant position-indicating events and the relevant classical law of motion.) For \emph{most} objects that are commonly called ``macroscopic'' such detectors do not exist. For \emph{all} so-called macroscopic objects this means that the probability of obtaining factual evidence of departures from their classically predictable motion is low.%
\footnote{This is where decoherence investigations~\cite{Zurek2003, Blanchardetal2000, Joosetal2003, Schlosshauer07}  come in. They demonstrate quantitatively that, where macroscopic objects are concerned, the probability of obtaining factual evidence of departures from ``classicality'' is extremely low.}
Hence there are \emph{many} so-called macroscopic objects of which the following is true: every one of their measured positions is consistent with every prediction that can be made on the basis of previous measurement outcomes and a classical law of motion. These are the objects that truly deserve to be labeled ``macroscopic.'' To permit a macroscopic pointer to indicate the value of an observable, one exception has to be made: its position can change unpredictably if and when it serves to indicate a value.

The macroworld can now be defined unambiguously as the totality of relative positions that exist between macroscopic objects---the totality of \emph{macroscopic positions}, for short. Macroscopic positions constitute the set of observables whose values can consistently be though of as factual \emph{per se}.

The crucial role played by the incompleteness of the world's spatiotemporal differentiation in demonstrating the consistency of attributing unconditional factuality to the macroworld deserves to be stressed. By definition, macroscopic positions never evince their indefiniteness (in the only way they could, through departures from classically predictable values). Macroscopic objects therefore follow trajectories that are only counterfactually ``fuzzy'': their positions are ``smeared out'' only in relation to an imaginary spatiotemporal background that is more differentiated than is the actual physical world. In the actual physical world there is no spatiotemporal background other than that defined by the positions of macroscopic objects and the times indicated by macroscopic clocks. Relative to this background, the positions of macroscopic objects and the times indicated by macroscopic clocks are definite as a matter of course. Hence we may, if we wish, attribute to the macroworld not only the unconditional factuality of the world of classical physics but also a genuine, albeit nonclassical, definiteness: macroscopic position are definite relative to the spatiotemporal background that actually exists in the physical world---they are as definite as it gets in the physical world. 

\section{Discussion}
The basic question is not how the ``unitary dynamics'' can be consistent with a definite macroworld. The basic question is: how can quantum mechanics be at once a fundamental physical theory and a ``mere'' probability calculus? As a probability calculus, it presupposes the value-indicating facts to which (and on the basis of which) it serves to assign probabilities. As a fundamental physical theory, it must make it possible to identify a set of observables whose values can consistently be though of as factual \emph{per se}, and this not just FAPP. 

The relative positions that exist between macroscopic objects constitute such a set. What makes it possible to demonstrate the strict consistency of attributing unconditional factuality to their values is that the spatiotemporal differentiation of the physical world is incomplete; it does not go ``all the way down.'' And what makes it possible to demonstrate this incompleteness is the interpretative principle proposed in Sec.~\ref{sec:yaip}.

Whenever quantum mechanics instructs us to first square the absolute values of amplitudes associated with alternatives and then add the result, there either is a matter of fact about the alternative taken, in which case we are dealing with ignorance probabilities, or the possibility of determining the alternative taken exists.%
\footnote{This possibility exists if the alternatives are correlated with those of a second system, in which case the alternative taken can be determined by a measurement on the second system.}
On the other hand, whenever quantum mechanics requires us to first add the amplitudes and then square the absolute value of the result, there cannot be any matter of fact about the alternative taken, and in this case we are not dealing with subjective or epistemic probabilities. We are dealing with probabilities that are objective (as opposed to subjective or epistemic). And the reason why they are objective (in this sense) is that the distinctions we make between the alternatives correspond to nothing in the real world. They exist solely in our minds.

In a way, Bub is right. The identification of an observable whose value can be consistently regarded as definite calls for a judicious choice on our part. I agree with Bub that the best (and arguably the only possible) choice is to stipulate the definiteness of the macroworld and of the instrument pointers it contains. However, this definiteness cannot be the definiteness of a classical position, which is sharp (dispersion-free) relative to an intrinsically and completely differentiated spatiotemporal background. It can only be the nonclassical definiteness of a macroscopic position, which is only counterfactually ``fuzzy'' (that is, ``smeared out'' only in relation to an imaginary spatiotemporal background that is more differentiated than is the actual physical world). What stands in the way of reconciling the entangled state ``at the end of a quantum measurement process'' with the definiteness of the macroscopic pointer reading is not the eigenvalue-eigenstate link, as Bub maintains, but a holdover from classical times---the belief that the spatiotemporal differentiation of the physical world goes ``all the way down.''

Moreover, the incompleteness of the world's spatiotemporal differentiation, which makes it possible to consistently attribute  unconditional factuality to macroscopic positions, can be demonstrated without recourse to decoherence, and so can the consistency of attributing unconditional factuality to macroscopic positions. Decoherence investigations merely confirm the conclusion that for sufficiently large and/or massive systems the probability of obtaining factual evidence of departures from ``classicality'' is extremely low, which ensures the existence of objects that are macroscopic according to the definition given in Sec.~\ref{sec:macroworld}.

Finally, whereas Bub provides no reason for his choice of interpretative principle other than that it stipulates the obvious (the consistency of the quantum-mechanical probability calculus with the existence of the events to which it serves to assign probabilities), the interpretative principle proposed in Sec.~\ref{sec:yaip} explains why quantum mechanics can be at once a fundamental physical theory and a ``mere'' probability calculus. In summary: it implies that the spatial distinctions we make cannot be intrinsic to space, that regions ``of space'' must be realized by macroscopic objects, that the spatiotemporal differentiation of the physical world is incomplete, that macroscopic positions are definite in a nonclassical sense, and that unconditional factuality can be consistently attributed to them.

\end{document}